\documentclass[a4paper,11pt,reqno,english]{amsart}
\usepackage[left=25mm,top=30mm,bottom=25mm,right=25mm]{geometry}
\usepackage{amsmath,amsfonts,dsfont,bm} %
\usepackage[utf8]{inputenc}     
\usepackage{cmap}               
\usepackage{hyperxmp}
\usepackage[english]{babel}
\usepackage[
  pdfdisplaydoctitle=true,
  colorlinks=true,
  urlcolor=blue,
  citecolor=blue,
  linkcolor=blue,
  pdfstartview=FitH,
  pdfpagemode=UseNone]{hyperref}

\usepackage{todonotes}

\newtheorem{thm}{Theorem}[section]

\theoremstyle{definition}

\theoremstyle{remark}
\newtheorem{rem}[thm]{Remark}
\newtheorem{exa}[thm]{Example}
\numberwithin{equation}{section}

\newcommand{\RR}{\mathbb{R}}                

\newcommand{\norm}[1]{\left\Vert#1\right\Vert}
\newcommand{\abs}[1]{\left\vert#1\right\vert}
\newcommand{\set}[1]{\left\{#1\right\}}

\newcommand{\GL}{\mathrm{GL}}               
\newcommand{\OO}{\mathrm{O}}                
\newcommand{\SO}{\mathrm{SO}}               
\newcommand{\DD}{\mathbb{D}}                
\newcommand{\octa}{\mathbb{O}}                

\newcommand{\so}{\mathfrak{so}}

\newcommand{\Ela}{\mathbb{E}\mathrm{la}}    
\newcommand{\TT}{\mathbb{T}}                
\newcommand{\Sym}{\mathbb{S}}               
\newcommand{\Orb}{\mathrm{Orb}}
\newcommand{\SC}{\Sigma}                    

\newcommand{\ee}{\bm{e}}                
\newcommand{\nn}{\bm{n}}                

\newcommand{\id}{\mathbf{1}}
\newcommand{\ba}{\mathbf{a}}
\newcommand{\bb}{\mathbf{b}}
\newcommand{\bh}{\mathbf{h}}
\newcommand{\bt}{\mathbf{t}}
\newcommand{\bu}{\mathbf{u}}
\newcommand{\bv}{\mathbf{v}}

\newcommand{\bI}{\mathbf{I}}                
\newcommand{\bC}{\mathbf{C}}                
\newcommand{\bE}{\mathbf{E}}
\newcommand{\bA}{\mathbf{A}}                
\newcommand{\bH}{\mathbf{H}}                
\newcommand{\bK}{\mathbf{K}}                
\newcommand{\bT}{\mathbf{T}}                
\newcommand{\bS}{\mathbf{S}}
\newcommand{\bR}{\mathbf{R}}

\DeclareMathOperator*{\argmin}{argmin}
\DeclareMathOperator{\tr}{tr}
\DeclareMathOperator{\fix}{Fix}
\DeclareMathOperator{\diag}{diag}

\hypersetup{
  pdfauthor={A. Antonelli, B. Desmorat, B. Kolev and R. Desmorat} 
  pdftitle={Distance to plane elasticity orthotropy by Euler-Lagrange method}, 
  pdfsubject={}, 
  pdfkeywords={Anisotropy; Distance to a symmetry class; orthotropic plane elasticity tensor}, 
  pdflang=en, 
  }

\begin{document}

\title[Distance to plane elasticity orthotropy]{Distance to plane  elasticity orthotropy \protect\\ by Euler-Lagrange method}

\author{A. Antonelli}
\address[Adrien Antonelli]{Université Paris-Saclay, ENS Paris-Saclay, CNRS,  LMT - Laboratoire de Mécanique et Technologie, 91190, Gif-sur-Yvette, France.}
\email{adrien.antonelli@ens-paris-saclay.fr}

\author{B. Desmorat}
\address[Boris Desmorat]{Sorbonne Université, UMPC Univ Paris 06, CNRS, UMR 7190, Institut d'Alembert, F-75252 Paris Cedex 05, France \& Univ Paris Sud 11, F-91405 Orsay, France}
\email{boris.desmorat@sorbonne-universite.fr}

\author{B. Kolev}
\address[Boris Kolev]{Université Paris-Saclay, ENS Paris-Saclay, CNRS,  LMT - Laboratoire de Mécanique et Technologie, 91190, Gif-sur-Yvette, France}
\email{boris.kolev@math.cnrs.fr}

\author{R. Desmorat}
\address[Rodrigue Desmorat]{Université Paris-Saclay, ENS Paris-Saclay, CNRS,  LMT - Laboratoire de Mécanique et Technologie, 91190, Gif-sur-Yvette, France}
\email{rodrigue.desmorat@ens-paris-saclay.fr}


\date{\today}%


\begin{abstract}
  Constitutive tensors are of common use in mechanics of materials. To determine the relevant symmetry class of an experimental tensor is still a tedious problem. For instance, it requires numerical methods in three-dimensional elasticity. We address here the more affordable case of plane (2D) elasticity, which has not been fully solved yet. We recall first Vianello's orthogonal projection method, valid for both the isotropic and the square symmetric (tetragonal) symmetry classes. We then solve in a closed-form the problem of the distance to plane elasticity orthotropy, thanks to the Euler-Lagrange method.
\end{abstract}

\maketitle

\section{Introduction}
\label{sec:intro}

Experimental techniques have been developed in order to measure all the components of an elasticity tensor, such as the ultrasonic measurements~\cite{KRG1971,AHR1991,Art1993,FBG1996,Del2005,BAA2006}. Once its components $C^{\mathrm{raw}}_{ijkl}$ are identified (in a working basis), one may wonder what is its relevant material symmetry (\emph{i.e.} symmetry group). For single crystals or for composite materials, a material symmetry is expected (for example cubic symmetry or orthotropy) so that the question becomes to find a tensor $\bC$, with a given material symmetry, the closest to the measured tensor $\bC^{\mathrm{raw}}$ (usually with no material symmetry at all, \emph{i.e.} triclinic in 3D, biclinic in 2D).

The precise identification of the symmetry class -- or the symmetry group -- of a constitutive tensor is not an easy task, mainly for two reasons \cite{FBG1996}:
\begin{itemize}
  \item the tensor is measured in a given orientation, possibly experimenter dependent, which may not coincide with an expected symmetry group, not allowing then for the recognition by eye (on the Kelvin matrix representation) of the orthotropic, tetragonal, cubic, \ldots well-known expressions for $\bC^{\mathrm{raw}}$ (the well-known normal forms of elasticity tensors \cite{Fed1968}),
  \item the experimental discrepancy/errors makes the material symmetries approximate.
\end{itemize}

Sufficient conditions (in \cite{BBS2007}) and necessary and sufficient conditions (in \cite{OKDD2021}) have been formulated to characterize, in an arbitrarily oriented coordinate system, the symmetry class of a three-dimensional elasticity tensor. For plane (2D) elasticity, the symmetry classes characterization is much simpler than in 3D~\cite{BOR1996,Via1997,VV2001}. Methods furthermore exist to bring back a rotated elasticity in its normal form \cite{Bae1998,ZTL2013,ADD2020}. Unfortunately, they are \emph{a priori} useless in the common case of a triclinic/biclinic measured elasticity tensor. When experimental discrepancy has to be dealt with, the literature approaches are based on the concept of distance of a tensor to a considered symmetry class~\cite{GTT1963,Fed1968,Fra1995,FBG1996,Hel1996,FGB1998,MN2006,KS2009,DKS2011}, starting from a given (usually measured) elasticity tensor $\bC^{\mathrm{raw}}$ with no material symmetry, and sometimes from the additional quantification of the measurement errors~\cite{Bon2009,DKS2015,GS2010}.

The 3D case has, by far, been the most studied. It remains the most challenging case, and determining the distance to a 3D symmetry class is usually done numerically, with the risk of reaching a local minimum instead of the global minimum~\cite{FGB1998,KS2009,DKS2015,MN2006}. Recently, in \cite{SMB2020}, following \cite{Bae1993}, the authors have used the basis carried by a second-order deviatoric tensor, a linear covariant $\bt(\bC^{asym})$, built from the Backus asymmetric part $\bC^{asym}=\bC^{\text{raw}}-(\bC^{\text{raw}})^{s}$ of the raw elasticity tensor. They obtain, this way, closed-form expressions for upper bounds estimates of the distances to the elasticity symmetry classes. In 2D, the asymmetric part $\bC^{asym}$ is always isotropic (see remark \ref{rem:isotropicA}) so that, in order to apply this procedure, another (non isotropic) second-order covariant of $\bC^{\text{raw}}$ has to be used.

The 2D case has been shown to be more affordable than the 3D case \cite{Via1997}. Indeed:
\begin{itemize}
  \item There are only four symmetry classes, easily characterized (instead of eight in three-dimensional elasticity \cite{FV1996});
  \item Both the isotropic and the tetragonal (square symmetry) strata\footnote{A symmetry stratum is the set of all tensors having the same symmetry class.} are linear subspaces of the vector space of plane elasticity tensors $\Ela(2)$. Closed form-expressions for the distance to these symmetry strata are obtained using orthogonal projections on these subspaces.
\end{itemize}

The problem of calculating the distance to plane elasticity orthotropy has also been posed in \cite{Via1997}, but not solved explicitly.
An upper bound estimate, expressed in terms of invariants, of this distance has been obtained in \cite{OLDK2021}.
It is the aim of the present work to derive a closed form solution for the exact 2D distance problem. For the sake of self-completeness, we will recall Vianello's \emph{orthogonal projection method} for the isotropic and the square symmetric cases (\autoref{sec:distance-stratum}). We will determine, by the \emph{Euler-Lagrange method}, the closest orthotropic plane elasticity tensor to a given raw tensor $\bC^{\mathrm{raw}}\in \Ela(2)$. The method is detailed in \autoref{sec:Euler-Lagrange} in a sufficiently general manner so that it also applies to other situations. For instance, the case of the distance to a transversely isotropic second order (3D) constitutive tensor (such as a thermal expansion or a thermal conductivity tensor) is treated in \autoref{sec:3D-S2}.
A discussion on different equivalent definitions of the distance to a symmetry class is provided in \autoref{sec:distance-defs}. In \autoref{sec:upper-bounds}, we give a proof that, concerning the distance to orthotropic plane elasticity, the upper bounds estimates proposed in \cite{SMB2020} and in \cite{OLDK2021} are equal.

Using the Euclidean structure of space, no distinction will be made between covariant, contravariant or mixed tensors. All tensor components will be expressed with respect to an orthonormal basis.

\section{Plane elasticity tensors}
\label{sec:planeC}

In this paper, we consider the natural action of the orthogonal group $\OO(2)$ on the vector space of elasticity tensors in two dimensions (sometimes, called plane elasticity tensors~\cite{Via1997,VV2001})
\begin{equation*}
  \Ela(2)=\set{\bC ;\; C_{ijkl}=C_{jikl}=C_{ijlk}=C_{klij}},
\end{equation*}
which describe constitutive equations for two-dimensional linearly elastic bodies. The orthogonal group $\OO(2)$ consists in two components. The subgroup, noted $\SO(2)$, of rotations $r_{\theta}$,
\begin{equation}\label{eq:rotations}
  r_{\theta} :=
  \begin{pmatrix}
    \cos \theta & - \sin \theta \\
    \sin \theta & \cos \theta
  \end{pmatrix},
\end{equation}
and the set (which is not a subgroup), noted $\SO(2)s$, consisting of axial symmetries with respect to the axis with angle $\theta/2$ with $Ox$. These reflections are written as the composition
\begin{equation}\label{eq:symmetries}
  r_{\theta} s =
  \begin{pmatrix}
    \cos \theta & \sin \theta  \\
    \sin \theta & -\cos \theta
  \end{pmatrix}, \quad \text{where} \quad s :=
  \begin{pmatrix}
    1 & 0  \\
    0 & -1
  \end{pmatrix},
\end{equation}
($s$ is the symmetry with respect to the $x$-axis) and the identity element is denoted by
\begin{equation*}
  e :=
  \begin{pmatrix}
    1 & 0 \\
    0 & 1
  \end{pmatrix}.
\end{equation*}
The action of the full orthogonal group $\OO(2)$ on the vector space $\Ela(2)$ is then written as
\begin{equation*}
  (\rho(g) \bC)_{ijkl} = g_{ip} g_{jq} g_{kr} g_{ls} C_{pqrs}, \qquad g \in \OO(2),\, \bC \in \Ela(2).
\end{equation*}

\subsection{Plane elasticity symmetry classes}
\label{S:elastclasses}

The vector space of plane elasticity tensors splits into four symmetry strata $\Sigma_{[H]}$~\cite{VV1986,BOR1996,Via1997,Van2005}, a symmetry stratum being the set of all tensors having the same symmetry class. They are indexed by the conjugacy class $[H]$ of some symmetry group $H$ (see~\autoref{sec:group-theory}) and are naturally ordered as follows
\begin{equation}\label{eq:symmetry-classes-order}
  \text{biclinic} \quad \longrightarrow \quad \text{orthotropic} \quad \longrightarrow \quad \text{square symmetric} \quad \longrightarrow \quad \text{isotropic}
\end{equation}
In the following we will refer to a symmetry stratum by its mechanical name, setting $\Sigma_{\mathrm{ortho}}$, $\Sigma_{\mathrm{square}}$ and $\Sigma_{\mathrm{iso}}$, and generically $\SC$, instead of the notation $\Sigma_{[H]}$ (which uses the definitions of subgroups of the orthogonal group $\OO(2)$).

Raw plane elasticity tensors are in general biclinic but one may expect, for several mechanical/material science reasons, a particular symmetry, the orthotropic, the square symmetry (also called tetragonal in~\cite{Via1997}) or the isotropic symmetry. A natural question is then:
\emph{how far is a given raw elasticity tensor $\bC^{\mathrm{raw}}\in \Ela(2)$ from an orthotropic, a square symmetric or an isotropic tensor?} In the following, non biclinic plane elasticity tensors, which are either orthotropic, square symmetric or isotropic, will be said to be \emph{at least orthotropic} and the set of all tensors which are at least orthotropic will be called the \emph{closed orthotropic stratum} and noted by $\overline{\Sigma}_{\mathrm{ortho}}$. Hence
\begin{equation*}
  \overline{\Sigma}_{\mathrm{ortho}} = \Sigma_{\mathrm{iso}} \cup  \Sigma_{\mathrm{square}} \cup  \Sigma_{\mathrm{ortho}}.
\end{equation*}
Similarly, \emph{at least square symmetric} elasticity tensors are either square symmetric or isotropic, and belong to the closed symmetry stratum
\begin{equation*}
  \overline{\Sigma}_{\mathrm{square}}= \Sigma_{\mathrm{iso}} \cup  \Sigma_{\mathrm{square}}.
\end{equation*}
The arrows in~\eqref{eq:symmetry-classes-order} represent this ``at least of a given symmetry'' ordering.

\subsection{Normal forms}

It is well-known that any symmetric second-order tensor $\ba$ can be rotated onto a diagonal tensor $\bA$. This diagonal representation is called a \emph{normal form} for $\ba$ under the action of the rotation group. Similarly, any plane elasticity tensor $\bC$ (which is at least orthotropic) can be written as $\bC=\rho(g)\bA$ where $\bA\in \Ela(2)$ has the following Kelvin matrix representation
\begin{equation}\label{eq:A}
  \underline \bA=\begin{pmatrix}
    A_{1111} & A_{1122} & 0          \\
    A_{1122} & A_{2222} & 0          \\
    0        & 0        & 2 A_{1212}
  \end{pmatrix}.
\end{equation}
This normal form depends on four independent parameters if it is orthotropic, three parameters if it has square symmetry ($A_{2222}=A_{1111}$), and two parameters if it is isotropic ($A_{222}=A_{1111}$ and $2A_{1212}=A_{1111}-A_{1122}$). We denote by $N$ this number of independent parameters.

\begin{rem}\label{rem:FixD2}
  Observe that in~\eqref{eq:A}, the tensor $\bA$ is fixed by the dihedral group $\DD_{2}:=\set{e,r_{\pi}, s, r_{\pi}s}$. Hence, the space of tensors $\bA$ having Kelvin matrix representation~\eqref{eq:A}, is noted $\fix(\DD_{2})$. A systematic way to calculate such a normal form for a tensor in a given symmetry class is described in~\autoref{sec:group-theory}.
\end{rem}

\subsection{Harmonic decomposition}
\label{S:HarmDecomp}

Following \cite{BOR1996,Via1997}, we denote by $\bC=(\lambda, \mu, \bh, \bH)$ the following explicit harmonic decomposition of a plane
elasticity tensor $\bC$, into two scalars $\lambda, \mu$, a second-order deviatoric (harmonic) tensor $\bh$ and a fourth order harmonic tensor $\bH$:
\begin{equation}
  \label{eq:HarmDecomp}
  \bC :=  2 \mu\, \bI + \lambda\, \id \otimes \id + \frac{1}{2} (\id \otimes \bh + \bh \otimes \id) + \bH ,
\end{equation}
where $\bI$ is the fourth order identity tensor with components $I_{ijkl}=\frac{1}{2} (\delta_{ik} \delta_{jl} + \delta_{il} \delta_{jk})$.
Explicitly, one has
\begin{align}
  \label{eq:lambdadeC}
  \lambda & :=\frac{1}{8} (3 C_{iijj} -2 C_{ijij}) = \frac{1}{8}(C_{1111} + 6 C_{1122} - 4 C_{1212} + C_{2222}),
  \\
  \mu     & := \frac{1}{8} (2C_{ijij} - C_{iijj}) = \frac{1}{8} (C_{1111}-2 C_{1122} + 4 C_{1212} + C_{2222}),
  \\
  \bh     & :=(\tr_{12} \bC)'
  \qquad
  \left(\textit{i.e.} \; h_{ij}=C_{kkij} -\frac{1}{2} C_{kkll} \delta_{ij}\right),
  \\
  \label{eq:HdeC}
  \bH     & := \bC- \frac{1}{2} (\id \otimes \bh + \bh \otimes \id) - 2 \mu\, \bI - \lambda \id \otimes \id,
\end{align}
where $(\cdot)'$ means the deviatoric part.
Recall that an \emph{harmonic tensor} is a totally symmetric and traceless tensor.

The two scalars $\lambda$ and $\mu$ and
\begin{align*}
   & I_{2}(\bC):= \norm{\bh}^{2}=\frac{1}{2}(C_{1111}-C_{2222})^{2} + 2 (C_{1112} + {C_{2212}})^{2},
  \\
   & J_{2}(\bC):= \norm{\bH}^{2}=\frac{1}{8} \left(C_{1111}-2 C_{1122}-4 C_{1212} + C_{2222}\right){}^{2} + 2 \left(C_{1112}-C_{1222}\right){}^{2},
\end{align*}
are invariants of the elasticity tensor.

Let $(\cdot)^{s}$ be the total symmetrization and $\odot$ be the (totally) symmetric tensor product. For symmetric second-order tensors $\ba, \bb$, we have
\begin{equation*}
  \ba \odot \bb = (\ba \otimes \bb)^s,
  \qquad
  (\ba \odot \bb)_{ijkl}=\frac{1}{6}(a_{ij} b_{kl}+a_{kl}b_{ij} +a_{ik} b_{jl}+b_{ik}a_{jl} +a_{il} b_{jk}+b_{il}a_{jk} ),
\end{equation*}

The explicit harmonic decomposition~\eqref{eq:HarmDecomp} can be rewritten as
\begin{equation}
  \bC =  \frac{2}{3}(\mu-\lambda) (\bI-\id \otimes \id)+(2 \mu+\lambda) \id \odot \id  + \frac{1}{2} (\id \odot \bh + \bh \odot \id) + \bH,
\end{equation}
thanks to the two following properties,
\begin{align}
  2 \, \bI + \id \otimes \id        & =  3 \, \id \odot \id,           &
  \id \otimes \bh + \bh \otimes \id & = \id \odot \bh + \bh \odot \id,
\end{align}
the second being specific to the 2D case when $\bh=\bh'$ is harmonic \cite{DD2016}.
The totally symmetric part $\bS$ and the asymmetric  part $\bC^{asym}$ --in the sense of Backus \cite{Bac1970}-- of a plane elasticity tensor $\bC$ are then
\begin{align*}
  \bS =\bC^{s}        & =(2 \mu+\lambda) \id \odot \id +\frac{1}{2} (\id \odot \bh + \bh \odot \id)+ \bH,
  \\
  \bC^{asym} =\bC-\bS & = \frac{2}{3}(\mu-\lambda) \left(\bI-\id \otimes \id \right),
\end{align*}

\begin{rem}\label{rem:isotropicA}
  We deduce from the last equality that, \emph{in 2D, the asymmetric part $\bC^{asym}$ of a plane elasticity tensor is always isotropic}.
\end{rem}

\section{Closest tensor in a given plane elasticity symmetry class}
\label{sec:distance-stratum}

Given an experimental raw tensor $\bC^{\mathrm{raw}}$, the aim is to find a plane elasticity tensor $\bC^{\SC}$ in a given (exact) symmetry stratum $\SC$ which is the closest to $\bC^{\mathrm{raw}}$, \emph{i.e.}, to calculate
\begin{equation}\label{eq:Copt1}
  \bC^{\SC} = \argmin_{\bC \in \SC} \norm{\bC^{\mathrm{raw}}-\bC},
\end{equation}
a problem which has already been extensively studied~\cite{GTT1963,Fra1995,FBG1996,Via1997,FGB1998}. Here, we choose to work with the Euclidean norm $\norm{\bC} =\sqrt{\langle \bC,\bC \rangle}$ (see~\cite{MN2006,MGD2019} for other norms), derived from the $\OO(2)$-invariant scalar product
\begin{equation*}
  \langle \bC^{1}, \bC^{2} \rangle = \bC^{1} :: \bC^{2} = C^{1}_{ijkl} C^{2}_{ijkl}.
\end{equation*}
Once a tensor $\bC^{\SC}$ solution of~\eqref{eq:Copt1} is found, the distance to the considered symmetry class $\SC$ is just
\begin{equation*}
  d(\bC^{\mathrm{raw}},\SC)=\norm{\bC^{\mathrm{raw}}-\bC^{\SC}}.
\end{equation*}
In general, several methods allow to solve such a minimization problem~\cite{FBG1996,KS2009,DKS2015,MN2006}, most of them appealing to numerical minimization schemes.

However, concerning the problem for the isotropic and the square symmetry plane elasticity symmetry classes, observe that both of the corresponding closed strata \emph{are linear subspaces} of $\Ela(2)$, which are described respectively by the following linear equations
\begin{itemize}
  \item $\bh = 0$ and $\bH=0$ for $\overline{\Sigma}_{\mathrm{iso}} = \Sigma_{\mathrm{iso}}$,
  \item $\bh = 0$ for $\overline{\Sigma}_{\mathrm{square}} =\Sigma_{\mathrm{iso}}\cup \Sigma_{\mathrm{square}}$,
\end{itemize}
where $\bh$ and $\bH$ are the second and fourth order components of the harmonic decomposition $\bC=(\lambda, \mu, \bh, \bH)$.
The distances to these symmetry strata are then obtained using orthogonal projections on these subspaces and no further calculations are required. The fact that the isotropic stratum is a linear subspace is very general but the fact that square symmetry stratum is a linear subspace is exceptional (symmetry strata are not in general linear spaces \cite{AS1983,AKP2014})
and seems to have been noticed first by Vianello~\cite{Via1997}. For the sake of self-completeness, we recall with full details these analytical solutions below, which are based on the harmonic decomposition of elasticity tensors.

\subsection{Closest isotropic tensor}

The isotropic elasticity tensor $\bC^{\mathrm{iso}}$ the closest to $\bC^{\mathrm{raw}}$ is the orthogonal projection on the isotropic stratum
\begin{equation}\label{eq:Ciso}
  \bC^{\mathrm{iso}} = 2 \mu\, \bI + \lambda \id \otimes \id,
\end{equation}
where the Lamé constants of $\bC^{\mathrm{iso}}$ are determined from $\bC^{\mathrm{raw}}$, as
\begin{equation*}
  \begin{aligned}
     & \lambda :=\frac{1}{8} (3 C^{\mathrm{raw}}_{iijj} -2 C^{\mathrm{raw}}_{ijij}) = \frac{1}{8}(C^{\mathrm{raw}}_{1111} + 6 C^{\mathrm{raw}}_{1122} - 4 C^{\mathrm{raw}}_{1212} + C^{\mathrm{raw}}_{2222}),
    \\
     & \mu := \frac{1}{8} (2C^{\mathrm{raw}}_{ijij} - C^{\mathrm{raw}}_{iijj}) = \frac{1}{8} (C^{\mathrm{raw}}_{1111}-2 C^{\mathrm{raw}}_{1122} + 4 C^{\mathrm{raw}}_{1212} + C^{\mathrm{raw}}_{2222}).
  \end{aligned}
\end{equation*}
This has allowed Vianello to obtain a nice expression for the distance of $\bC^{\mathrm{raw}}$ to isotropy
\begin{equation*}
  d(\bC^{\mathrm{raw}}, \Sigma_{\mathrm{iso}})=\sqrt{I_{2}(\bC^{\mathrm{raw}}) + J_{2}(\bC^{\mathrm{raw}})}
\end{equation*}

\subsection{Closest square symmetric tensor}

In the same manner, the at least square symmetric (tetragonal) elasticity tensor $\bC^{\mathrm{square}}$ the closest to $\bC^{\mathrm{raw}}$ is the orthogonal projection of $\bC^{\mathrm{raw}}$ on the square symmetry stratum
\begin{equation}
  \bC^{\mathrm{square}}=\bC^{\mathrm{iso}}+ \bH,
\end{equation}
with both $\bC^{\mathrm{iso}}$ (by~\eqref{eq:Ciso}) and $\bH$ (by~\eqref{eq:lambdadeC}-\eqref{eq:HdeC}) determined from $\bC^{\mathrm{raw}}$. If $\bH\neq 0$, $\bC^{\mathrm{square}}$ is square symmetric.
The Kelvin matrix representation of the square symmetric elasticity tensor which is the closest to $\bC^{\mathrm{raw}}$ is then
\begin{equation*}
  \underline \bC^{\mathrm{square}}=\left(
  \begin{array}{cccccc}
      \frac{1}{2}(C^{\mathrm{raw}}_{1111} + C^{\mathrm{raw}}_{2222})      & C^{\mathrm{raw}}_{1122}                                             & \frac{\sqrt{2}}{2}(C^{\mathrm{raw}}_{1112}-C^{\mathrm{raw}}_{2212}) \\
      C^{\mathrm{raw}}_{1122}                                             & \frac{1}{2}(C^{\mathrm{raw}}_{1111} + C^{\mathrm{raw}}_{2222})      & \frac{\sqrt{2}}{2}(C^{\mathrm{raw}}_{2212}-C^{\mathrm{raw}}_{1112}) \\
      \frac{\sqrt{2}}{2}(C^{\mathrm{raw}}_{1112}-C^{\mathrm{raw}}_{2212}) & \frac{\sqrt{2}}{2}(C^{\mathrm{raw}}_{2212}-C^{\mathrm{raw}}_{1112}) & 2 C^{\mathrm{raw}}_{1212}                                           \\
    \end{array}
  \right).
\end{equation*}
and the distance of $\bC^{\mathrm{raw}}$ to square (tetragonal) symmetry, obtained by Vianello, is the nice formula
\begin{equation*}
  d(\bC, \Sigma_{\mathrm{square}})=\sqrt{I_{2}(\bC^{\mathrm{raw}})}.
\end{equation*}

\section{Closest orthotropic tensor using Euler-Lagrange method}
\label{sec:Euler-Lagrange}

Starting from a raw (usually experimental) plane elasticity tensor $\bC^{\mathrm{raw}}\in \Ela(2)$, assumed biclinic, it remains to determine the closest orthotopic tensor $\bC^{\mathrm{ortho}}$ to $\bC^{\mathrm{raw}}$. Since the orthotropic stratum is not a linear subspace, a different approach is required. In what follows, we will apply the \emph{Euler-Lagrange method}, using the parametrization $\bC = \rho(g)\bA$ \cite{Via1997,Del2005}, where $g \in \OO(2)$ and $\underline \bA$ is the normal form~\eqref{eq:A}.

\begin{rem}\label{rem:redundent-solutions}
  The parametrization $\bC = \rho(g)\bA$ is not one to one, because several orthogonal transformations $g$ may correspond to the same tensor $\bC=\rho(g)\bA$, depending on the symmetry group of $\bA$. In particular, redundant solutions are obtained as $\bC=\rho(gh)\bA$, when $h$ is in the symmetry group of $\bA$.
\end{rem}

Since the reflection $s$ belongs to the symmetry group of $\bA$, namely the dihedral group $\DD_{2}$, we deduce that for each solution $(g, \bA)$ of the optimization problem, it corresponds a redundant solution $(gs, \bA)$. Therefore, we can reduce the minimization problem to the set $\SO(2) \times \fix(\DD_{2})$, rather than $\OO(2) \times \fix(\DD_{2})$, where $\fix(\DD_{2})$ denotes the subspace of normal forms for at least orthotropic elasticity tensors. We shall thus consider the optimization problem
\begin{equation}\label{eq:Copt2}
  d(\bC^{\mathrm{raw}}, \Sigma_{\mathrm{ortho}}) = \min_{g, \bA} \norm{\bC^{\mathrm{raw}} - \rho(g) \bA }
\end{equation}
over $g=r_{\theta} \in \SO(2)$ and $\bA \in \fix(\DD_{2})$ (with $N=4$ independent components $A_{ijkl}$).

Solving problem~\eqref{eq:Copt2} can be done numerically with the risk to reach a local minimum instead of the global minimum~\cite{FGB1998,KS2009,DKS2015,MN2006}. To avoid this difficulty in 3D, François and coworkers did propose pole figures for the given elasticity tensor $\bC^{\mathrm{raw}}$\cite{Fra1995,FBG1996} (renamed plots of the monoclinic distance in~\cite{KS2009,MN2006}). Accordingly, they got an initial value for $\bC=\rho(g) \bA$, not too far from $\bC^{\mathrm{raw}}$, which was then optimized by a standard numerical (iterative)  scheme.
In the following, one will avoid numerical schemes and a closed form solution will be sought.

\subsection{Euler-Lagrange first order equations}

Since the method is very general, we explain it below for a linear action $\rho(g)\bT$ of the orthogonal group $\SO(d)$ (where $d$ is $2$ or $3$) on a space $\TT$ of tensors $\bT$. We assume that we know a normal form $\bA$ for a given symmetry class (with $N$ independent parameters). Moreover, we use a scalar product on tensors $\bT$, for which the action is isometric, meaning that
\begin{equation}\label{eq:isometric-action}
  \langle \rho(g)\bT_{1}, \rho(g)\bT_{2} \rangle = \langle \bT_{1}, \bT_{2} \rangle,
\end{equation}
for all $g \in \SO(d)$ and all tensors $\bT_{1}$, $\bT_{2}$. The problem is now, given a raw tensor $\bT^{\mathrm{raw}}$, to calculate the critical points of the functional
\begin{equation}\label{eq:f}
  f(g,\bA) = \norm{\bT^{\mathrm{raw}}-\rho(g) \bA}^{2} = \langle \bT^{\mathrm{raw}} - \rho(g)\bA,\bT^{\mathrm{raw}} - \rho(g)\bA\rangle.
\end{equation}
To do so, we consider a path $g(t) \in \SO(d)$ with $g(0) = g$ and $\dot{g}(0) = \delta g$ and a path $\bA(t)$ of normal forms with $\bA(0) = \bA$ and $\dot{\bA}(0) = \delta \bA$, and we get
\begin{equation}\label{eq:f-differential}
  df(g,\bA).(\delta g, \delta \bA) = \left.\frac{d}{dt}\right|_{t=0} f(g(t),\bA(t)) = -2\langle \bT^{\mathrm{raw}}-\rho(g)\bA, (\delta\rho(g))\bA + \rho(g)\delta \bA \rangle.
\end{equation}

To calculate $\delta\rho(g)$, we use the infinitesimal action $\rho'$ induced by $\rho$ (see \autoref{sec:group-theory}) and defined by
\begin{equation}\label{eq:rhopu}
  \rho'(\bu)\bT = \left.\frac{d}{dt}\right|_{t=0} \rho(h(t))\bT,
\end{equation}
where $h(t)$ is a path in $\SO(d)$ with $h(0)=I$ and $\dot h(0) = \bu \in \so(d)$ is an infinitesimal rotation.

\begin{exa}
  For an elasticity tensor $\bT=\bC\in \Ela(d)$, we get
  \begin{equation}\label{eq:rhopuC}
    (\rho'(\bu)\bC)_{ijkl} = u_{ip} C_{pjkl} + u_{jp} C_{ipkl} + u_{kp} C_{ijpl} + u_{lp} C_{ijkp}.
  \end{equation}
\end{exa}

\begin{rem}
  Due to the fact that $\rho$ is isometric~\eqref{eq:isometric-action}, we have
  \begin{equation}\label{eq:antisymrhu}
    \langle \bT_{1},\rho'(\bu)\bT_{2}\rangle = -\langle\rho'(\bu)\bT_{1},\bT_{2} \rangle,
  \end{equation}
  for every infinitesimal rotation $\bu \in \so(d)$ and every tensors $\bT_{1}$, $\bT_{2}$.
\end{rem}

Now, introducing the infinitesimal rotations $\bv = g^{-1}\delta g$ and $\bu = \delta g\, g^{-1}$, and, thanks to the fact that $\rho$ is a linear representation, we have (see \autoref{sec:group-theory})
\begin{equation*}
  \delta\rho(g) = \rho(g)\rho'(\bv) = \rho'(\bu) \rho(g), \quad \text{where} \quad \bv = g^{-1}\bu\, g.
\end{equation*}

Therefore, the Euler-Lagrange first-order equations (deduced from~\eqref{eq:f-differential})
\begin{equation*}
  df(g,\bA).(\delta g, \delta \bA) =0, \qquad \forall \delta g, \delta \bA,
\end{equation*}
recast as
\begin{equation*}
  \langle \bT^{\mathrm{raw}}-\rho(g)\bA,\rho(g)\delta \bA \rangle = 0,  \qquad \forall \delta \bA,
\end{equation*}
and
\begin{equation*}
  \langle \bT^{\mathrm{raw}}-\rho(g) \bA,\rho'(\bu)\rho(g) \bA \rangle = 0 \qquad \forall \bu.
\end{equation*}
Moreover, by~\eqref{eq:isometric-action}, the  first equation rewrites as
\begin{equation}\label{eq1}
  \langle \rho(g^{t})\bT^{\mathrm{raw}}- \bA, \delta \bA \rangle = 0,  \qquad \forall \delta \bA,
\end{equation}
and by~\eqref{eq:antisymrhu}, we have
\begin{equation*}
  \langle \rho(g)\bA,\rho'(\bu)\rho(g)\bA \rangle = - \langle \rho'(\bu)\rho(g)\bA,\rho(g)\bA \rangle = 0
\end{equation*}
so that the second equation simplifies as
\begin{equation}\label{eq:Interm}
  \langle \bT^{\mathrm{raw}},\rho'(\bu)\rho(g) \bA \rangle = 0, \qquad \forall \bu.
\end{equation}

Consider now a basis $(\bu^\alpha)$ of the space of skew symmetric matrices (of dimension $d(d-1)/2$) and an orthonormal basis $(\bE^{I})$ of the space of tensors, such that  the first $N$ vectors $\bE^{1}, \dotsc , \bE^N$ span the subspace of normal forms $\bA$ and set
\begin{equation}\label{eq:decomp}
  \bA = \sum_{I=1}^N \Lambda_{I}\, \bE^{I} .
\end{equation}
Then, the first order equations~\eqref{eq1}-\eqref{eq:Interm} can be written directly in terms of the unknown variables, the rotation $g = (g_{ij})$ and the components $\Lambda_{I}$ of $\bA$, as the two sets of equations
\begin{align}\label{eq:O11}
   & \Lambda_{I} = <\rho(g^{t})\bT^{\mathrm{raw}},\bE^{I}>                                   &  & 1 \le I \le N,
  \\
  \label{eq:O12}
   & <\rho(g^{t})\bT^{\mathrm{raw}}, \rho'(\bu^\alpha)\sum_{I=1}^N  \Lambda_{I} \bE^{I}> = 0 &  & 1 \le \alpha \le d(d-1)/2.
\end{align}
These equations are polynomial in the variables $g_{ij}$ and $\Lambda_{I}$. Using the genericity of $\bT^{\mathrm{raw}}$ and some arguments of algebraic geometry, one can prove that they have only a finite number of solutions $(g_{k}, \bA^{k})$ when the symmetry group corresponding to the symmetry stratum under consideration is finite. These solutions are the critical points of the function $f$ introduced in~\eqref{eq:f}. Therefore, only a finite number of critical points $(g_{k}, \bA^{k})$ needs to be compared. The global minimum of $\norm{\bT^{\mathrm{raw}}-\bT}$ with $\bT$ of a given symmetry class $\SC$, will simply be
\begin{equation*}
  d(\bT^{\mathrm{raw}}, \SC)=\min_{k} \norm{\bT^{\mathrm{raw}}-\bT^{k}},
  \qquad
  \bT^{k}= \rho(g_{k}) \bA^{k},
\end{equation*}
and the closest tensor(s) to the considered symmetry class will be
\begin{equation*}
  \bT^{\SC}=\argmin_{k} \norm{\bT^{\mathrm{raw}}-\bT^{k}}.
\end{equation*}

\subsection{Distance to plane elasticity orthotropy}
\label{sec:2D-elasticity}

We now apply the preceding method to calculate the distance to the orthotropic stratum of a given plane elasticity tensor $\bC^{\mathrm{raw}}$. Three upper bounds estimates of this distance \cite{GTT1963,SMB2020,SMB2020} are compared in \autoref{sec:upper-bounds}.

The Lie algebra $\so(2)$ (the space of infinitesimal 2D rotations) of the rotation group $\SO(2)$ is the one-dimensional vector space spanned by
\begin{equation*}
  \bu^{1} = \frac{1}{\sqrt{2}}
  \begin{pmatrix}
    0 & -1 \\
    1 & 0
  \end{pmatrix}.
\end{equation*}

It was shown in~\cite{Via1997} that $\bC$ and $\bC^{\mathrm{raw}}$ have the same isotropic part $\bC^{\mathrm{iso}}$ defined by~\eqref{eq:Ciso}. Therefore, according to~\eqref{eq:A}, the parametrization of $\bC^{\mathrm{ortho}}$ can be written as
\begin{equation}\label{eq:param-Ela}
  \bC^{\mathrm{ortho}}=\bC^{\mathrm{iso}} + \bT, \qquad \bT= \rho(g)(\Lambda_{1} \bE^{1} + \Lambda_{2} \bE^{2}),
  \qquad
  g= \begin{pmatrix}
    \cos \theta & - \sin \theta \\
    \sin \theta & \cos \theta
  \end{pmatrix},
\end{equation}
where the orthonormal basis of tensors $(\bE^{1},\bE^{2})$ have Kelvin matrix representations
\begin{equation*}
  \underline \bE^{1}=\left(
  \begin{array}{ccc}
      \frac{1}{2 \sqrt{2}}  & -\frac{1}{2 \sqrt{2}} & 0                   \\
      -\frac{1}{2 \sqrt{2}} & \frac{1}{2 \sqrt{2}}  & 0                   \\
      0                     & 0                     & -\frac{1}{\sqrt{2}} \\
    \end{array}
  \right),
  \qquad
  \underline \bE^{2} =\left(
  \begin{array}{ccc}
      \frac{1}{\sqrt{2}} & 0                   & 0 \\
      0                  & -\frac{1}{\sqrt{2}} & 0 \\
      0                  & 0                   & 0 \\
    \end{array}
  \right) ,
\end{equation*}
with $\langle \bC^{\mathrm{iso}},\bE^{1}\rangle = \langle\bC^{\mathrm{iso}},\bE^{2}\rangle = 0$. The later equalities can be checked
using the Kelvin matrix representation of $\bC^{\mathrm{iso}}$,
\begin{equation*}
  \underline \bC^{\mathrm{iso}}=\left(
  \begin{array}{ccc}
      2 \mu + \lambda & \lambda         & 0     \\
      \lambda         & 2 \mu + \lambda & 0     \\
      0               & 0               & 2 \mu \\
    \end{array}
  \right).
\end{equation*}

We are then looking for the critical points of the functional ~\eqref{eq:f} with $\bT^{\mathrm{raw}}=\bC^{\mathrm{raw}}-\bC^{\mathrm{iso}}$ given and where $\bA=\Lambda_{1} \bE^{1} + \Lambda_{2} \bE^{2}$ is bi-dimensional. Equation~\eqref{eq:O11} is used to determine the components $\Lambda_{1}$ and $\Lambda_{2}$ as functions of the angle of the rotation~\eqref{eq:rotations}, giving
\begin{equation}\label{eq:Lambdai-Ela}
  \Lambda_{1}(\theta)=X_{1} \cos 4 \theta + Y_{1} \sin 4 \theta
  ,\qquad
  \Lambda_{2}(\theta)=X_{2} \cos 2 \theta + Y_{2} \sin 2 \theta ,
\end{equation}
with
\begin{align*}
   & X_{1}=\frac{1}{2 \sqrt{2}} (C^{\mathrm{raw}}_{1111}-2 C^{\mathrm{raw}}_{1122}-4 C^{\mathrm{raw}}_{1212} + C^{\mathrm{raw}}_{2222}),
  \\
   & Y_{1}=\sqrt{2} (C^{\mathrm{raw}}_{1112}-C^{\mathrm{raw}}_{2212})  ,
  \\
   & X_{2}=\frac{1}{\sqrt{2}}(C^{\mathrm{raw}}_{1111}-C^{\mathrm{raw}}_{2222}) ,
  \\
   & Y_{2}=\sqrt{2} (C^{\mathrm{raw}}_{1112} + C^{\mathrm{raw}}_{2212}) .
\end{align*}
Then, these results are injected in Equation~\eqref{eq:O12}, using~\eqref{eq:rhopuC}, in order to determine $\theta $. This gives
\begin{equation}\label{eq:theta-Ela}
  A \cos 8 \theta + B \sin 8 \theta + C \cos 4 \theta + D \sin 4 \theta =0 ,
\end{equation}
with the following closed form expressions for the constants
\begin{equation*}
  A =4 X_{1} Y_{1} ,
  \qquad
  B = 2 (Y_{1}^{2}-X_{1}^{2}),
  \qquad
  C=2 X_{2} Y_{2} ,
  \qquad
  D=Y_{2}^{2}-X_{2}^{2} .
\end{equation*}

The sought tensors $\bC^{k}$ are finally given by~\eqref{eq:param-Ela} with $\theta_{k}$ solution of~\eqref{eq:theta-Ela}.

\begin{rem}
  Equation~\eqref{eq:theta-Ela} has several solutions $\theta$ (some corresponding to maxima, others to saddle points, and other to minima, possibly local). By remark~\ref{rem:redundent-solutions}, if $(\theta,\bA)$ is a solution, then $(\theta + \pi,\bA)$ is also a solution. Therefore, to obtain all the solutions of the Euler-Lagrange equations, it is sufficient to seek for $\theta$ in an interval of length $\pi$. Furthermore, if $\theta$ is a solution of~\eqref{eq:theta-Ela}, then $\theta-\frac{\pi}{2}$ is also a solution of~\eqref{eq:theta-Ela} (with identical $\Lambda_{1}$ but with opposite $\Lambda_{2}$). This allows us to seek for $\theta$ in an interval of length $\pi/2$, but to take account, for each solution $\theta_{k} \in ]-\frac{\pi}{4}, \frac{\pi}{4}[$, of a twin solution $\theta_{-k} = \theta_{k}-\frac{\pi}{2}\in ]-\frac{3\pi}{4}, -\frac{\pi}{4}[$.
\end{rem}

Check first if $\theta_{1}=\pi/4$ and  $\theta_{-1}=-\pi/4$ are solutions (an improbable event in case of an experimental, therefore noisy, elasticity tensor $\bC^{\mathrm{raw}}$). Seek for the other solutions as $\theta = \frac{1}{2} \arctan t$, where $t$ is a real root of the degree 4 polynomial (generically, $A-C \ne 0$)
\begin{equation*}
  (A-C) t^{4} + (2 D-4 B) t^{3} -6 A t^{2} + (2 D + 4 B)t + A + C=0,
\end{equation*}
deduced from~\eqref{eq:theta-Ela} when $\theta\neq \frac{\pi}{4} + n\frac{ \pi}{2}$.

Finally, choosing among the solutions $\bC^{k}$, corresponding to solutions $\theta_{k}$ and their twins $\theta_{-k}=\theta_{k} - \frac{\pi}{2}$, the plane elasticity tensor
\begin{equation*}
  \bC^{\mathrm{ortho}}=\bC^{\ell}=\bC^{\mathrm{iso}} + \rho(g(\theta_\ell))\left(\Lambda_{1}(\theta_\ell) \bE^{1} + \Lambda_{2}(\theta_\ell) \bE^{2}\right),
\end{equation*}
which realizes the global minimum
\begin{equation*}
  d(\bC^{\mathrm{raw}}, \Sigma_{\mathrm{ortho}}) = \min_{k} \norm{\bC^{\mathrm{raw}} - \bC^{k}} = \norm{\bC^{\mathrm{raw}} - \bC^{\mathrm{ortho}}},
\end{equation*}
completes the solving.

\section{Different equivalent definitions of the distance to a symmetry class}
\label{sec:distance-defs}

We have considered (and solved in a closed form for plane elasticity) the \emph{distance to a symmetry class problem} formulated thanks to the normal form $\bA$ / rotation $g$ parameterization (Eq.~\eqref{eq:Copt2}), as  \cite{Via1997,DVS1998,Del2005}
\begin{equation}\label{eq:Copt2math}
  d(\bC^{\mathrm{raw}}, \Sigma_{[H]}) := \min_{g\in \SO(d),\, \bA \in \fix H} \norm{\bC^{\mathrm{raw}} - \rho(g) \bA },
  \qquad
  \bC^{\mathrm{raw}}, \bA \in \Ela(d),
\end{equation}
with $\Sigma_{[H]}=\Sigma_{[\DD_{2}]}=\Sigma_\mathrm{ortho}$ the orthotropic stratum.

\begin{rem}
  Here, the notation $[H]$ stands for the conjugacy class of $H$, that is the set of all subgroups $gHg^{-1}$, where $g\in \SO(d)$ (see \autoref{sec:group-theory}). Indeed, a symmetry class is described by all the symmetry groups $gHg^{-1}$ rather than the subgroup $H$ itself. For instance, the orthotropy is described by the conjugacy class $[\DD_{2}]$, where the dihedral group $H=\DD_{2}$ defined in remark~\ref{rem:FixD2} is just a representative in this class. The set $\Sigma_{[H]}$ correspond to all tensors having the same symmetry class $[H]$ and for this reason $\Sigma_{[gHg^{-1}]} = \Sigma_{[H]}$.
\end{rem}

Alternative definitions of $d(\bC^{\mathrm{raw}}, \Sigma_{[H]})$ introduce the \emph{Reynolds operator} $\bR_H: \Ela(d) \to \fix(H)$  associated here with the finite group\footnote{For infinite compact groups $H$ such as $\OO(2)$ or $\SO(2)$, the finite sum is replaced by an integral and the Haar measure, a (bi-invariant) probability measure on $H$~\cite{Ste1994}, is used.} $H$ \cite[Chapter 2]{Stu2008},
\begin{equation}\label{eq:RH}
  \bR_H\left(\bC\right):=\frac{1}{\abs{H}}\sum_{h\in H} \rho(h)\bC,
\end{equation}
where $\abs{H}$ is the order of the group $H$.     The Reynolds operator is an orthogonal projector, meaning that, for all tensors $\bC$,
\begin{equation}\label{eq:ReyOrtho}
  \bR_{H}\left(\bR_{H}(\bC)\right) = \bR_{H}\left(\bC\right), \quad\text{and}\quad \left\langle \bC-\bR_{H}\left(\bC\right), \bR_{H}\left(\bC\right) \right\rangle=0.
\end{equation}
For the conjugate subgroup $gHg^{-1}$ of $H$, we get
\begin{equation}\label{eq:RgHgm1}
  \bR_{gHg^{-1}}\left(\bC\right)=\frac{1}{\abs{H}}\sum_{h\in H} \rho(ghg^{-1})\bC
  =\left(\rho(g)\bR_H \rho(g^{-1})\right)\left(\bC\right) .
\end{equation}
In particular, for plane elasticity and $H=\DD_{2}$ (of order $\abs{\DD_{2}}=4$), we have
\begin{align*}
   & \bR_{\DD_{2}}\left(\bC\right)=\frac{1}{4}\left(\bC+\rho(r_{\pi})\bC+ \rho(s)\bC +\rho(r_{\pi} s)\bC\right)
  =\frac{1}{2}\left(\bC+ \rho(s)\bC \right),
  \\
   & \bR_{g\DD_{2} g^{-1}}\left(\bC\right)=\frac{1}{2}\left(\bC+ \rho(g s g^{-1})\bC \right),
\end{align*}
where the elements $e$, $r_{\pi}$, $s$ and $r_{\pi}s$ of the dihedral group $\DD_{2}$ have been defined in \autoref{sec:planeC}.

\vskip 2mm

Another definition of the distance to a symmetry class is due to François and coworkers \cite{Fra1995,FGB1998},
\begin{equation}\label{eq:CoptReynolds}
  d(\bC^{\mathrm{raw}}, \Sigma_{[H]}) := \min_{g\in \SO(d)} \norm{\bC^{\mathrm{raw}} - \bR_{gHg^{-1}} \left(\bC^{\mathrm{raw}}\right) }.
\end{equation}
and a slightly modified definition of it has recently been used by Weber et al \cite{WGB2019},
\begin{equation}\label{eq:defequivdist}
  d(\bC^{\mathrm{raw}}, \Sigma_{[H]})
  = \min_{g\in \SO(d)} \norm{\rho(g)\bC^{\mathrm{raw}} - \bR_{H}\left((\rho(g) \bC^{\mathrm{raw}}\right) }.
\end{equation}

\begin{rem}
  The three distances  to a symmetry class defined by~\eqref{eq:Copt2math},~\eqref{eq:CoptReynolds} and~\eqref{eq:defequivdist} are equal when the norm $\norm{\cdot}$ is $\SO(d)$-invariant (meaning that $\norm{\rho(g) \bC}=\norm{\bC}$ for all rotations $g$ and tensors $\bC$), which is the case for the Euclidean norm.
\end{rem}

We provide now a proof of this fact. Consider first the definition~\eqref{eq:Copt2math},
\begin{equation*}
  d(\bC^{\mathrm{raw}}, \Sigma_{[H]})=\min_{g\in \SO(d)} \min_{\bA\in \fix{H}} \norm{\bC^{\mathrm{raw}} - \rho(g) \bA },
\end{equation*}
because the variables $g$ and $\bA$ are independent. We have then
\begin{align*}
  \min_{\bA\in \fix{H}} \norm{\bC^{\mathrm{raw}} - \rho(g) \bA }=\min_{\bA\in \fix{H}} \norm{\rho(g)^{-1}\bC^{\mathrm{raw}} - \bA},
\end{align*}
and hence, by~\eqref{eq:ReyOrtho}, the minimizer $\bA\in \fix{H}$ is the orthogonal projection
\begin{equation*}
  \bA= \bR_{H}\left(\rho(g)^{-1}\bC^{\mathrm{raw}}\right)
\end{equation*}
of $\rho(g)^{-1}\bC^{\mathrm{raw}}$ on $\fix{H}$. Thus, using~\eqref{eq:RgHgm1}, we get
\begin{equation*}
  \min_{\bA\in \fix{H}} \norm{\bC^{\mathrm{raw}} - \rho(g) \bA }=\norm{\rho(g)^{-1}\bC^{\mathrm{raw}} -  \bR_{H}\left(\rho(g)^{-1}\bC^{\mathrm{raw}}\right) }
  =\norm{\bC^{\mathrm{raw}} - \bR_{gHg^{-1}} \left(\bC^{\mathrm{raw}}\right) }
  ,
\end{equation*}
which proves the equivalence of definitions~\eqref{eq:Copt2math} and~\eqref{eq:CoptReynolds}. Now, since $\rho(g)$ preserves the norm of elasticity tensors, we have moreover
\begin{align*}
  d(\bC^{\mathrm{raw}}, \Sigma_{[H]})
  =\min_{g\in \SO(d)}\norm{\rho(g^{-1})\bC^{\mathrm{raw}} -  \bR_{H}\left(\rho(g^{-1})\bC^{\mathrm{raw}}\right) }
  =\min_{g\in \SO(d)}\norm{\rho(g)\bC^{\mathrm{raw}} -  \bR_{H}\left(\rho(g)\bC^{\mathrm{raw}}\right) },
\end{align*}
which shows the equivalence with definition~\eqref{eq:defequivdist}.

\section{Conclusion}

The problem of calculating the distance of a raw tensor to a symmetry class has been posed. It has been fully solved for plane (bidimensional) elasticity tensors. First, Vianello's orthogonal projection method, valid for both the isotropic and the square symmetric (tetragonal) symmetry classes, has been recalled. Then, the remaining case of the distance to plane elasticity orthotropy has been solved, thanks to Euler-Lagrange method. The solution proposed is analytical, it requires only to find the roots of a degree four polynomial and to compare the at most eight closed-form solutions of the first-order Euler-Lagrange equations. The method is general and relies on the use of the infinitesimal action of Lie algebra of the rotation group $\SO(d)$ to solve the first-order equations. This use seems to be new in the present context. Another important feature of this method is that, dealing with polynomial functions, there is generically only a finite number of critical points, provided that the symmetry group associated to the symmetry class under consideration is finite. In that case, finding the global minimum is immediate. To illustrate further this method, an application to constitutive (3D) symmetric second-order tensors is provided in \autoref{sec:3D-S2}.

\appendix

\section{Basic concepts in representation theory}
\label{sec:group-theory}

A linear action of a group $G$ on a vector space $V$ (also usually called a \emph{linear representation} of $G$ on $V$) is a mapping
\begin{equation*}
  \rho : G \to \GL(V), \qquad g \mapsto \rho(g).
\end{equation*}
where $\rho(g)$ is an invertible linear transformation of $V$ and $\GL(V)$ is the group of invertible, linear mappings of $V$ into itself, and such that
\begin{equation*}
  \rho(e) = \mathrm{Id}, \quad \text{and} \quad \rho(g_{1} g_{2}) = \rho(g_{1}) \rho(g_{2}),
\end{equation*}
where $g_{1}, g_{2} \in G$ and $e$ is the unit element of $G$. The \emph{orbit} of a vector $v\in V$ is the set
\begin{equation*}
  \Orb(v) := \set{\rho(g) v;\; g\in G}.
\end{equation*}
The \emph{symmetry group} of a vector $v \in V$ is the subgroup of $G$ defined by
\begin{equation*}
  G_{v} := \set{g\in G;\; \rho(g) v = v},
\end{equation*}
and its \emph{symmetry class}, noted $[G_{v}]$, is defined as the conjugacy class of $G_{v}$ in $G$, \textit{i.e.},
\begin{equation*}
  [G_{v}] := \set{gG_{v}g^{-1};\; g\in G}.
\end{equation*}
Observe that all vectors in a same orbit $\Orb(v)$ have the same symmetry class, since
\begin{equation*}
  G_{\rho(g) v} = gG_{v}g^{-1}.
\end{equation*}

Finally, to each symmetry class $[H]$, corresponds a \emph{symmetry stratum} $\Sigma_{[H]}$ (which, in general, is not a linear subspace of $V$) which is the set of all vectors $v$ which have exactly the symmetry $[H]$
\begin{equation*}
  \Sigma_{[H]} := \set{v\in V;\; G_{v}\in [H]}.
\end{equation*}

\begin{rem}
  To each symmetry stratum $\Sigma_{[H]}$ corresponds a \emph{closed symmetry stratum} $\overline{\Sigma}_{[H]}$, which is the set of vectors $v\in V$ which have \emph{at least} the symmetry $[H]$.
\end{rem}

Given a symmetry class $[H]$ and choosing a representative $H$ in this conjugacy class, we can build the \emph{fix point set}
\begin{equation*}
  \fix(H) := \set{v \in V;\;\rho(h) v = v; \; \forall h \in H}.
\end{equation*}
This set is a linear subspace of $V$ and intersects each orbit $\Orb(v)$ when $H \subset G_{v}$. The subspace $\fix(H)$ defines a \emph{normal form} for $\overline{\Sigma}_{[H]}$, if the representative $H$ in $[H]$ has been well chosen.

We will suppose now that $G$ is a \emph{Lie group} (which means that $G$ is not only a group but also a differentiable manifold and that the group operations are smooth). In practice, all groups $G$ considered are closed subgroup of the general linear group (and hence $G$ can be considered as a matrix group). Then, we can assume that the representation $\rho$ of $G$ on $V$ is smooth. In that case, it induces a linear mapping, called the \emph{infinitesimal action}
\begin{equation*}
  \rho^{\prime} : \mathfrak{g} \to \mathfrak{gl}(V), \qquad \bu \mapsto T_{e}\rho.\bu,
\end{equation*}
where $\mathfrak{g} = T_{e}G$, the tangent space at the identity element is called the \emph{Lie algebra} of $G$. This infinitesimal action is sufficient to describe the linear tangent mapping at every point $g$ since the relation $\rho(g_{1}g_{2}) = \rho(g_{1})\rho(g_{2})$ leads to
\begin{equation*}
  T_{g}\rho.\delta g = \rho(g) \rho^{\prime}(g^{-1}\,\delta g) = \rho^{\prime}(\delta g \, g^{-1}) \rho(g) .
\end{equation*}

\section{3D symmetric second-order tensors}
\label{sec:3D-S2}

In this section, Euler-Lagrange method, described in \autoref{sec:Euler-Lagrange}, is illustrated on the space $\TT=\Sym^{2}(\RR^{3})$ of three-dimensional symmetric second-order tensors. We consider a given (for example experimental) constitutive symmetric second-order tensor $\bT^{\mathrm{raw}}=\ba^{\mathrm{raw}}$ in 3D (for example an anisotropic thermal expansion tensor or an anisotropic conductivity tensor), which is furthermore assumed to have three distinct eigenvalues (\emph{i.e.} to be orthotropic).

The action of the rotation group $\SO(3)$ on
\begin{equation*}
  \TT = \Sym^{2}(\RR^{3}) = \set{\ba ;\; a_{ji}=a_{ij}},
\end{equation*}
is written $\rho(g) \ba=g\, \ba\, g^{t}$ ($g \in \SO(3)$), or in components
\begin{equation*}
  (\rho(g) \ba)_{ij}= g_{ip} g_{jq} a_{pq}.
\end{equation*}
Note that, on even order tensors, the action of the full orthogonal group $\OO(3)$ cannot be distinguished from the action of the 3D rotation group $\SO(3)$.

The space of three-dimensional symmetric second-order tensors splits into three symmetry classes: orthotropy (three distinct eigenvalues), transverse isotropy (two distinct eigenvalues) and isotropy (one single eigenvalue). Since a generic symmetric second-order tensor $\bT^{\mathrm{raw}}=\ba^{\mathrm{raw}}$ is orthotropic, a natural question then arise of \emph{how far it is from a transversely isotropic or an isotropic tensor ?}

The distance to isotropy is obtained straightforwardly using the orthogonal projection of $\ba^{\mathrm{raw}}$ onto the space of spherical tensors
\begin{equation*}
  \ba^{\mathrm{iso}}= \frac{1}{3} (\tr \ba^{\mathrm{raw}}) \id
\end{equation*}
and the distance of $\ba^{\mathrm{raw}}$ to isotropy is
\begin{equation*}
  d(\ba^{\mathrm{raw}}, \textrm{isotropy})=\norm{\ba^{\mathrm{raw}}- \ba^{\mathrm{iso}}}=\sqrt{\ba^{\mathrm{raw}\,\prime}:\ba^{\mathrm{raw}\,\prime}}.
\end{equation*}

We will thus focus on the distance of $\ba^{\mathrm{raw}}$ to transverse isotropy, and seek by the Euler-Lagrange method for the closest transversely isotropic symmetric second-order tensor $\ba$ to $\ba^{\mathrm{raw}}$ (assumed orthotropic). The parametrization of (at least) transversely isotropic tensors is given by
\begin{equation*}
  \ba = \rho(g) \bA = g \bA g^{t},
\end{equation*}
where $g \in \SO(3)$ and $\bA=\Lambda_{1} \bE^{1} + \Lambda_{2} \bE^{2}$, with
\begin{equation*}
  \bE^{1}=\frac{1}{\sqrt{6}}
  \begin{pmatrix}
    1 & 0 & 0  \\
    0 & 1 & 0  \\
    0 & 0 & -2
  \end{pmatrix},
  \qquad
  \bE^{2}=\frac{1}{\sqrt{3}}
  \begin{pmatrix}
    1 & 0 & 0 \\
    0 & 1 & 0 \\
    0 & 0 & 1
  \end{pmatrix}.
\end{equation*}
This choice corresponds to an axis $\nn=\ee_{3}$ of transverse isotropy for $\bA$.

\begin{rem}
  Note that, contrary to the case of the distance to orthotropy for plane elasticity tensors, the symmetry group involved here for transverse isotropy is not finite. We expect thus to find an infinity of solutions $(g,\bA)$ but they will lead, anyway, to a finite number of tensors $\ba = g \bA g^{t}$, candidate to be a global minimum for the distance to the transversely isotropic stratum.
\end{rem}

The problem is to determine the minimum of the functional
\begin{equation*}
  f(g, \bA) = \norm{\ba^{\mathrm{raw}}- g \bA g^{t}}^{2},
  \qquad
  \bA=\Lambda_{1} \bE^{1} + \Lambda_{2} \bE^{2}, \quad g \in \SO(3).
\end{equation*}
Without loss of generality, we can assume (after a diagonalization of $\ba^{\mathrm{raw}}$) that the proper basis of $\ba^{\mathrm{raw}}$ is the canonical basis of $\RR^{3}$, $(\ee_{i})$, which will be done in the sequel.

The Lie algebra $\so(3)$ of the rotation group $\SO(3)$ is the vector space of $3 \times 3$ skew symmetric matrices (infinitesimal rotations), with as basis
\begin{equation*}
  \bu^{1}=\frac{1}{\sqrt{2}}
  \begin{pmatrix}
    0 & -1 & 0 \\
    1 & 0  & 0 \\
    0 & 0  & 0
  \end{pmatrix}
  , \quad
  \bu^{2}=\frac{1}{\sqrt{2}}
  \begin{pmatrix}
    0 & 0 & -1 \\
    0 & 0 & 0  \\
    1 & 0 & 0
  \end{pmatrix}
  , \quad
  \bu^{3}= \frac{1}{\sqrt{2}}
  \begin{pmatrix}
    0 & 0 & 0  \\
    0 & 0 & -1 \\
    0 & 1 & 0
  \end{pmatrix} .
\end{equation*}

To solve the problem, we calculate first the infinitesimal action $\rho'$ on $\Sym^{2}(\RR^{3})$, which is written as
\begin{equation*}
  (\rho'(\bu)\ba)_{ij}= u_{ip} a_{pj} + u_{jp} a_{ip},
\end{equation*}
or, in a more intrinsic form, as
\begin{equation}\label{eq:rhopu-order2}
  \rho'(\bu)\ba= \bu \ba - \ba \bu=[\bu, \ba].
\end{equation}

Then, we recast~\eqref{eq:Interm} using this expression, which leads to
\begin{align*}
  0 = \langle \ba^{\mathrm{raw}},\rho'(\bu)\rho(g) \bA \rangle & = \tr \left(\ba^{\mathrm{raw}} \big(\bu(\rho(g) \bA) - (\rho(g) \bA) \bu\big) \right)
  \\
                                                               & = \tr \left( \bu  \big(\rho(g) \bA \, \ba^{\mathrm{raw}}-  \ba^{\mathrm{raw}} \rho(g) \bA \big) \right)
  \\
                                                               & = \bu : \left[\rho(g) \bA,  \ba^{\mathrm{raw}} \right]
  \\
                                                               & =
  \bu :  \left[\ba,  \ba^{\mathrm{raw}} \right].
\end{align*}
Since this last equality is true for every skew symmetric matrix $\bu$, it implies that the commutator $\left[\ba,  \ba^{\mathrm{raw}} \right]$ vanishes and thus that the symmetric second-order tensors $\ba$ and $\ba^{\mathrm{raw}}$ commute. The given tensor $\ba^{\mathrm{raw}}$ being orthotropic, it has three distinct eigenvalues and $\left[\ba, \ba^{\mathrm{raw}} \right]=0$ means that $\ba$ is diagonal in the basis $(\ee_{i})$. Since $\ba = g \bA g^{t}$, either both $\ba$ and $\bA$ are isotropic, in which case, all rotations $g \in \SO(3)$ are solutions or both $\ba$ and $\bA$ are transversely isotropic, in which case, each solution $g$ of $\ba = g \bA g^{t}$ must send the transverse isotropy axis of $\bA$ onto the one of $\ba$ and writes thus $g = g_{k}h$, where
\begin{equation*}
  g_{1} = R(\ee_{2},\pi/2), \qquad g_{2} = R(\ee_{1},-\pi/2), \qquad g_{3} = \id,
\end{equation*}
$R(\nn, \theta)$ denotes the 3D rotation of angle $\theta$ around the vector $\nn$ and $h$ belongs to the subgroup of $\SO(3)$ of rotations which do not change the $Oz$ axis. This subgroup of 3D rotations is isomorphic to the orthogonal group $\OO(2)$. We have moreover
\begin{equation*}
  \bA = \Lambda_{1} \bE^{1} + \Lambda_{2} \bE^{2},
\end{equation*}
where, by~\eqref{eq:O11}, for each rotation $g$ solution of $\ba = g \bA g^{t}$,
\begin{equation*}
  \Lambda_{1}=(g^{t}\ba^{\mathrm{raw}}g):\bE^{1}, \qquad \Lambda_{2}=\frac{1}{\sqrt{3}}\tr \ba^{\mathrm{raw}}.
\end{equation*}
This achieves the determination of the critical points of $f$.

\begin{rem}
  One recognizes $\Lambda_{2} \bE^{2}= \ba^{\mathrm{iso}}$ as the isotropic part of $\ba^{\mathrm{raw}}$.
\end{rem}

Isotropic solutions have already been calculated and correspond to the unique solution
\begin{equation*}
  \ba^{\mathrm{iso}}= \frac{1}{3} (\tr \ba^{\mathrm{raw}}) \id,
\end{equation*}
but which may not be a global minimum in the present situation (and, indeed, will not be one generically). Transversely isotropic solutions $(g, \bA)$ correspond finally to the three following candidates for the global minimum $\ba^{\mathrm{trans-iso}}$
\begin{equation*}
  \left(\begin{array}{ccc}
      \lambda_{1} & 0                                   & 0                                   \\
      0           & \frac{\lambda_{2} + \lambda_{3}}{2} & 0                                   \\
      0           & 0                                   & \frac{\lambda_{2} + \lambda_{3}}{2} \\
    \end{array}
  \right)
  \;\text{or}\;
  \left(\begin{array}{ccc}
      \frac{\lambda_{1} + \lambda_{3}}{2} & 0           & 0                                   \\
      0                                   & \lambda_{2} & 0                                   \\
      0                                   & 0           & \frac{\lambda_{1} + \lambda_{3}}{2} \\
    \end{array}
  \right)
  \;\text{or}\;
  \left(
  \begin{array}{ccc}
      \frac{\lambda_{1} + \lambda_{2}}{2} & 0                                   & 0           \\
      0                                   & \frac{\lambda_{1} + \lambda_{2}}{2} & 0           \\
      0                                   & 0                                   & \lambda_{3} \\
    \end{array}
  \right)
\end{equation*}
where $\lambda_{1} < \lambda_{2} < \lambda_{3}$ are the distinct eigenvalues of $\ba^{\mathrm{raw}}$. The global minimum is obtained by comparing the three candidate distances $\norm{\ba^{\mathrm{raw}}- \ba }$ associated with these critical points. We will summarize these results with the following conclusion.

Let $\ba^{\mathrm{raw}}=\diag(\lambda_{1},\lambda_{2},\lambda_{3}) \in \Sym^{2}(\RR^{3})$ be a given orthotropic symmetric second-order tensor, with $\lambda_{1} < \lambda_{2} <\lambda_{3}$. Then, the transversely isotropic second-order tensor
$\ba^{\mathrm{trans-iso}} \in \Sym^{2}(\RR^{3})$ closest to $\ba^{\mathrm{raw}}$ (for the Frobenius norm) commutes with $\ba^{\mathrm{raw}}$ and writes
\begin{align*}
          & \ba^{\mathrm{trans-iso}} =\left(
  \begin{array}{ccc}
      \frac{\lambda_{1} + \lambda_{2}}{2} & 0                                   & 0           \\
      0                                   & \frac{\lambda_{1} + \lambda_{2}}{2} & 0           \\
      0                                   & 0                                   & \lambda_{3} \\
    \end{array}
  \right)
          &                                  & \textrm{if} \quad \lambda_{2} < \frac{1}{2}(\lambda_{1} + \lambda_{3}),
  \\
          & \ba^{\mathrm{trans-iso}} =\left(
  \begin{array}{ccc}
      \lambda_{1} & 0                                   & 0                                   \\
      0           & \frac{\lambda_{2} + \lambda_{3}}{2} & 0                                   \\
      0           & 0                                   & \frac{\lambda_{2} + \lambda_{3}}{2} \\
    \end{array}
  \right) &                                  & \textrm{if} \quad \lambda_{2} > \frac{1}{2}(\lambda_{1} + \lambda_{3}).
\end{align*}
in the proper basis of $\ba^{\mathrm{raw}}$. The distance of $\ba^{\mathrm{raw}}$ to the transversely isotropic class is then
\begin{equation*}
  d(\ba^{\mathrm{raw}},\textrm{transverse isotropy}) = \frac{1}{\sqrt{2}} \min_{i \ne j} \abs{\lambda_{i}-\lambda_{j}} .
\end{equation*}

\section{Upper bounds estimates of the distance to a symmetry class}
\label{sec:upper-bounds}

Upper bounds estimates of the distance $d(\bC^{\mathrm{raw}}, \Sigma_{[H]})$ to a symmetry class have been proposed in the literature, for instance in \cite{GTT1963} and \cite{SMB2020} for 3D elasticity symmetry classes, and in \cite{OLDK2021} for 2D orthotropic symmetry class. They are easier to calculate than the distance itself.

First, Gazis and coworkers did simply suggest the estimate
\begin{equation}
  M_{g_{0}} (\bC^{\mathrm{raw}}, \Sigma_{[H]}) := \norm{\rho(g_{0})\bC^{\mathrm{raw}} - \bR_{H}\left(\rho(g_{0})\bC^{\mathrm{raw}}\right) },
\end{equation}
with $g_{0}$ a "well-chosen" rotation\footnote{in fact these authors did take $g_{0}=e$ the identity in their example.}, for example related to the manufacturing process or to the observed microstructure. Obviously,
\begin{equation}
  M_{g_{0}}(\bC^{\mathrm{raw}}, \Sigma_{[H]}) \geq  d(\bC^{\mathrm{raw}}, \Sigma_{[H]}),
\end{equation}
since the minimization in~\eqref{eq:defequivdist} is not performed when calculating $M_{g_{0}}(\bC^{\mathrm{raw}}, \Sigma_{[H]})$.

Studying the 3D elasticity problem, Stahn and coworkers astutely defined in \cite{SMB2020} a smaller upper bound $M(\bC^{\mathrm{raw}}, \Sigma_{[H]})$ of the distance to a 3D elasticity symmetry class, in a few steps,
\begin{enumerate}
  \item by computing
        \begin{equation*}
          \bt=\bC^{asym}:\id-\frac{1}{4}(\id:\bC^{asym}:\id)\, \id,
        \end{equation*}
        which is a second-order covariant of the asymmetric part $\bC^{asym}=\bC^{\mathrm{raw}}-(\bC^{\mathrm{raw}})^{s}$ of the given elasticity tensor $\bC^{\mathrm{raw}}$,

  \item following \cite{Bae1993}, by computing an eigenbasis of $\bt$ and a rotation $g_{0}$ that brings it into its diagonal form
        (\emph{i.e.}, such that $\rho(g_{0})\bt = g_{0} \, \bt\, g_{0}^{t}$ is diagonal),

  \item by computing $\rho(g_{0}) \bC^{\mathrm{raw}}$ (and assuming in fact that $\rho(g_{0}) \bC$ is close to the normal form of the sought tensor with higher symmetry),

  \item finally, by setting
        \begin{equation}\label{eq:MStahn}
          M(\bC^{\mathrm{raw}}, \Sigma_{[H]}): = \min_{g\in \octa}
          \norm{\rho(g)\rho(g_{0})\bC^{\mathrm{raw}} - \bR_{H}\left(\rho(g)\rho(g_{0})\bC^{\mathrm{raw}}\right) }
        \end{equation}
        with $\bR_H$ the Reynolds operator~\eqref{eq:RH}.
\end{enumerate}
The procedure introduces the proper cubic (octahedral) group $ \octa$
which is defined by
\begin{equation*}
  \octa = \set{g\in \SO(3);\; g \ee_{i} = \pm \ee_{j}},
\end{equation*}
where $(\ee_{i})$ is the canonical orthonormal basis of $\RR^{3}$. This group is of order $\abs{\octa}=24$. Generically, the second-order tensor $\bt$ is orthotropic, and, in 3D, there are exactly 24 rotations which bring $\bt$ into a diagonal form. Given such a rotation, say $g_{0}$, the other ones are written as
\begin{equation*}
  g g_{0} \quad\text{with} \quad g \in \octa,
\end{equation*}
so that definition~\eqref{eq:MStahn} does not depend on a particular choice for $g_{0}$.

\begin{rem}\label{rem:Majorant}
  As the $\min$ in~\eqref{eq:MStahn} is over the finite group $\octa$ (and not over the rotation group $\SO(3)$), one has
  \begin{align*}
    M(\bC^{\mathrm{raw}}, \Sigma_{[H]})
     & =
    \min_{g\in \octa}
    \norm{\rho(gg_{0})\bC^{\mathrm{raw}} - \bR_{H}\left(\rho(g g_{0})\bC^{\mathrm{raw}}\right) }
    \\
     &
    \geq \min_{g\in \SO(3)}
    \norm{\rho(g)\bC^{\mathrm{raw}} - \bR_{H}\left(\rho(g)\bC^{\mathrm{raw}}\right) }
    = d(\bC^{\mathrm{raw}}, \Sigma_{[H]}),
  \end{align*}
  as $g=gg_{0}$ is a rotation, and where the $\min$ over the subgroup $\octa\subset \SO(3)$ in the first line is generally strictly larger than the min over $\SO(3)$ in the last line. Hence, $M(\bC^{\mathrm{raw}}, \Sigma_{[H]})$ is an upper bound of the distance $d(\bC^{\mathrm{raw}}, \Sigma_{[H]})$.
\end{rem}

\begin{rem}\label{rem:isotropict}
  In 2D, the asymmetric part $\bC^{asym}$ of an elasticity tensor is isotropic (see remark \ref{rem:isotropicA}), so that the Stahn and coworkers' procedure  applies only if the definition of the second-order tensor $\bt$ is changed, for example into the second-order harmonic component $\bh$ of  $\bC^{\text{raw}}$ defined in section \ref{S:HarmDecomp},
  \begin{equation*}\label{eq:newt}
    \bt=\bh= (\tr_{12} \bC^{\text{raw}})'=(\id:\bC^{\text{raw}})'.
  \end{equation*}
  $M(\bC^{\mathrm{raw}}, \Sigma_{[H]})$ built from $\bt=\bh$ is then, by remark \ref{rem:Majorant}, an upper bound estimate of the distance $d(\bC^{\mathrm{raw}}, \Sigma_{[H]})$.
\end{rem}

Finally, an invariant formula has been obtained in 2D for an upper bound estimate of the distance to orthotropic elasticity \cite[Corollary 3.2]{OLDK2021}.
Indeed, setting
\begin{equation*}
  \bh:= (\tr_{12} \bC^{\text{raw}})'
  , \qquad
  I_{2}:=\norm{\bh}^{2}=\bh:\bh
  , \qquad
  J_{2}:=\norm{\bH}^{2}=\bH :: \bH
  , \qquad
  K_{3}:= \bh : \bH : \bh
  ,
\end{equation*}
where $\bH$ is furthermore the fourth-order harmonic part of $\bC^{\text{raw}}=(\lambda, \mu, \bh, \bH)$ (determined by~\eqref{eq:HdeC}),
and
\begin{equation*}
  \bC^{\text{upper}}:=  2 \mu\, \bI + \lambda \id \otimes \id
  + \frac{1}{2} (\id \otimes \bh + \bh \otimes \id)
  +\frac{2 K_{3}}{I_{2}^{2}}
  \left(\bh\odot \bh -\frac{1}{4} \norm{\bh}^{2} \id \odot \id \right),
\end{equation*}
then the positive invariant
\begin{equation}\label{eq:UpperNous}
  \Delta (\bC^{\mathrm{raw}}, \Sigma_{\text{ortho}}) := \norm{\bC^{\text{raw}}-\bC^{\text{upper}}} = \frac{\sqrt{J_{2} I_{2}^{2} - 2K_{3}^{2}}}{I_{2}},
\end{equation}
is an upper bound of the distance $d(\bC^{\mathrm{raw}}, \Sigma_{\text{ortho}})$ of the plane elasticity tensor $\bC^{\text{raw}}$ to orthotropy.

\begin{rem}
  In the particular case of orthotropic plane elasticity, using definition~\eqref{eq:newt} for $\bt$, we have the equality of the upper bounds estimates proposed in \cite{OLDK2021} and in \cite{SMB2020},
  \begin{equation*}
    \Delta (\bC^{\mathrm{raw}}, \Sigma_{\text{ortho}}) = \norm{\bC^{\text{raw}}-\bC^{\text{upper}}} = \frac{\sqrt{J_{2} I_{2}^{2} - 2 K_{3}^{2}}}{I_{2}} = M(\bC^{\mathrm{raw}}, \Sigma_{[H]}).
  \end{equation*}
\end{rem}

A proof of this result is as follows. The harmonic decomposition of the raw plane elasticity tensor is $\bC^{\text{raw}}=(\lambda, \mu, \bh, \bH)$ (by formulas~\eqref{eq:lambdadeC}--\eqref{eq:HdeC}). Let $g_{0}$ be a rotation such that
\begin{equation*}
  \bh_{0}:=\rho(g_{0})\bh=g_{0} \, \bh\, g_{0}^{t}
\end{equation*}
is diagonal. In 2D, there are only two such rotations, $g_{0}$ and $g_{1} = r_{\frac{\pi}{2}}g_{0}$, the second one giving $\rho(g_{1})\bh = -\bh_{0}$ (as $\bh$ is deviatoric). Since the harmonic decomposition is equivariant, we get
\begin{equation*}
  \rho(g_{0})\bC^{\text{raw}} = (\lambda, \mu, \bh_{0}, \bH_{0}).
\end{equation*}
where $\bH_{0}:= \rho(g_{0}) \bH$. The Stahn and coworkers upper bound is then (by definition~\eqref{eq:MStahn})
\begin{align*}
   & M(\bC^{\mathrm{raw}}, \Sigma_{\text{ortho}})
  \\
   & = \min \left( \norm{\rho(g_{0})\bC^{\mathrm{raw}} - \bR_{\DD_{2}}\left(\rho(g_{0})\bC^{\mathrm{raw}}\right)}, \norm{\rho(r_{\frac{\pi}{2}})\rho(g_{0})\bC^{\mathrm{raw}} - \bR_{\DD_{2}}\left(\rho(r_{\frac{\pi}{2}})\rho(g_{0})\bC^{\mathrm{raw}}\right)}
  \right)
  \\
   & = \min \left( \norm{\rho(g_{0})\bC^{\mathrm{raw}} - \bR_{\DD_{2}}\left(\rho(g_{0})\bC^{\mathrm{raw}}\right)} ,
  \norm{\rho(g_{0})\bC^{\mathrm{raw}} - \left(\rho(r_{\frac{\pi}{2}})^{-1}\bR_{\DD_{2}}\rho(r_{\frac{\pi}{2}})\right)\left(\rho(g_{0})\bC^{\mathrm{raw}}\right)}
  \right)
  \\
   & = \norm{\rho(g_{0})\bC^{\mathrm{raw}} - \bR_{\DD_{2}}\left(\rho(g_{0})\bC^{\mathrm{raw}}\right) }
  \\                                    & = \norm{\bH_{0}- \bR_{\DD_{2}}\left(\bH_{0}\right) },
\end{align*}
since $r_{\frac{\pi}{2}}\DD_{2}r_{\frac{\pi}{2}}^{-1} = \DD_{2}$ in 2D and thus
\begin{equation*}
  \rho(r_{\frac{\pi}{2}})^{-1}\,\bR_{\DD_{2}}\,\rho(r_{\frac{\pi}{2}}) = \bR_{r_{\frac{\pi}{2}}\DD_{2}r_{\frac{\pi}{2}}^{-1}} = \bR_{\DD_{2}}.
\end{equation*}
Using the harmonic decomposition of $\rho(g_{0})\bC^{\text{raw}}$, the upper bound~\eqref{eq:UpperNous} is
\begin{equation*}
  \Delta (\bC^{\mathrm{raw}}, \Sigma_{\text{ortho}}) = \Delta (\rho(g_{0}) \bC^{\mathrm{raw}}, \Sigma_{\text{ortho}})
  = \norm{\bH_{0}-2\frac{ \bh_{0}:\bH_{0}:\bh_{0}}{\norm{\bh_{0}}^{4}}
    \left(\bh_{0}\odot \bh_{0} -\frac{1}{4} \norm{\bh_{0}}^{2} \id \odot \id \right)}
\end{equation*}
and we are left to show that
\begin{equation*}
  2 \frac{\bh_{0}:\bH_{0}:\bh_{0}}{\norm{\bh_{0}}^{4}} \left(\bh_{0}\odot \bh_{0} - \frac{1}{4} \norm{\bh_{0}}^{2} \id \odot \id\right)
  = \bR_{\DD_{2}}\left(\bH_{0}\right),
\end{equation*}
for all $\bh_{0}\neq 0$ deviatoric and diagonal, and all fourth-order harmonic tensor $\bH_{0}$, which can be checked by comparing these expressions in components. Indeed, a general harmonic tensor $\bK$, has for matrix components
\begin{equation*}
  \underline \bK=
  \left(
  \begin{array}{ccc}
    K_{1111}          & -K_{1111}          & \sqrt{2} K_{1112}  \\
    -K_{1111}         & K_{1111}           & -\sqrt{2} K_{1112} \\
    \sqrt{2} K_{1112} & -\sqrt{2} K_{1112} & -2 K_{1111}        \\
  \end{array}
  \right).
\end{equation*}
Now set
\begin{equation*}
  \bT=2\frac{\bh_{0}:\bK:\bh_{0}}{\norm{\bh_{0}}^{4}}
  \left(\bh_{0}\odot \bh_{0} - \frac{1}{4} \norm{\bh_{0}}^{2} \id \odot \id \right),
\end{equation*}
where
\begin{equation*}
  \bh_{0}=\begin{pmatrix}
    \lambda & 0         \\
    0       & - \lambda
  \end{pmatrix},
\end{equation*}
with $\lambda\neq 0$ (otherwise, the elasticity tensor $\bC^{\text{raw}}$ has already the square symmetry). We get then
\begin{equation*}
  \underline \bT =
  \left(
  \begin{array}{ccc}
    K_{1111}  & -K_{1111} & 0           \\
    -K_{1111} & K_{1111}  & 0           \\
    0         & 0         & -2 K_{1111} \\
  \end{array}
  \right),
\end{equation*}
which is indeed the Kelvin matrix form of $\bR_{\DD_{2}}\left(\bK\right)$.

We have therefore shown that for a biclinic elasticity tensor $\bC^{\mathrm{raw}}\in \Ela(2)$
\begin{equation*}
  \Delta (\bC^{\mathrm{raw}}, \Sigma_{\text{ortho}})=M (\bC^{\mathrm{raw}}, \Sigma_{\text{ortho}})= \frac{\sqrt{J_{2} I_{2}^{2}-2 K_{3}^{2}}}{I_{2}}
  \geq d (\bC^{\mathrm{raw}}, \Sigma_{\text{ortho}}).
\end{equation*}

\subsection*{Funding} Two of the authors, R. Desmorat and B. Kolev, were partially supported by CNRS Projet 80--Prime GAMM (Géométrie algébrique complexe/réelle et mécanique des matériaux).


\end{document}